\documentclass{emulateapj} 
\usepackage{psfig}

\slugcomment{Accepted by the Astronomical Journal}

\shorttitle{Flux Densities and Polarization of Two Pulsars} 

\shortauthors{Crawford \& Tiffany}

\begin{document}

\title{Flux Densities and Radio Polarization Characteristics of Two
Vela-like Pulsars}

\author{Fronefield Crawford\altaffilmark{1,2} and Chelsea
L. Tiffany\altaffilmark{3}}

\altaffiltext{1}{Department of Physics and Astronomy, Franklin \&
Marshall College, P.O. Box 3003, Lancaster, PA 17604; email:
fcrawfor@fandm.edu}

\altaffiltext{2}{Department of Physics, Haverford College, Haverford,
PA 19041}

\altaffiltext{3}{Department of Astronomy, Wellesley College, 106
Central St., Wellesley, MA 02481}

\begin{abstract}
We report on dual-frequency radio polarimetry observations of two
young, energetic pulsars, PSRs J0940$-$5428 and J1301$-$6305.  These
were among the first Vela-like pulsars discovered in the Parkes
Multibeam survey.  We conducted observations of these pulsars with the
Australia Telescope Compact Array (ATCA) at center frequencies of 1384
and 2496 MHz using pulsar gating while preserving full Stokes
parameters.  After correcting for bandwidth depolarization, we have
measured polarization characteristics, flux densities, and rotation
measures for these pulsars.  The spectral indices derived from the
ATCA data are shallow but still consistent with values seen for
pulsars of this type.  The rotation measures for both pulsars are
consistent with those reported recently using data from the Parkes
telescope, and both pulsars have highly linearly polarized pulse
profiles at both 1384 and 2496 MHz.  Our results support a previously
noted correlation between high degree of linear polarization, shallow
spectral index, and large spin-down luminosity.
\end{abstract}


\section{Introduction} 

PSRs J0940$-$5428 and J1301$-$6305 were among the first pulsars
discovered in the Parkes Multibeam survey \citep{mlc+01}. Both pulsars
have fast spin periods ($P = 88$ and 185 ms, respectively), and both
are young, with characteristic ages $\tau_{c} \equiv P/2\dot{P}$ of 42
and 11 kyr, respectively. They also have large spin-down luminosities:
$\dot{E} \equiv 4 \pi^{2} I \dot{P} / P^{3} = 1.9 \times 10^{36}$ and
$1.7 \times 10^{36}$ erg s$^{-1}$, respectively \citep{mlc+01}, where
a moment of inertia of $I = 10^{45}$ g cm$^{2}$ is assumed. These
characteristics place them in the category of Vela-like pulsars, which
are generally defined as fast-spinning pulsars having characteristic
ages $10 \la \tau_{c} \la 100$~kyr and spin-down luminosities $\dot{E}
\ga 10^{36}$ erg s$^{-1}$ (e.g., Kramer et al. 2003\nocite{kbm+03}).

The radio polarization properties of such pulsars are useful to
measure for several reasons. Rotation measures (RMs) are used to probe
Galactic magnetic fields \citep{hml+06} and can support associations
between young pulsars and radio supernova remnants (e.g., Crawford \&
Keim 2003\nocite{ck03}; Caswell et al. 2004\nocite{cmc04}). Young,
energetic pulsars generally show a higher degree of linear
polarization than older and less energetic pulsars \citep{hkk98}, and
their polarization fractions and phase-resolved polarization
characteristics can be used to constrain the pulsar's emission
geometry (e.g., Lyne \& Manchester 1988; Everett \& Weisberg
2001\nocite{lm88,ew01}; Johnston \& Weisberg 2006\nocite{jw06}).  In a
number of cases, young pulsars have been observed to have
single-peaked pulse profiles that are wide and highly linearly
polarized (e.g., Crawford, Manchester, \& Kaspi 2001\nocite{cmk01}),
which may indicate emission from only one part of a wider conal beam
\citep{m96}.  Pulsars with these kinds of emission properties also
typically have shallow radio spectral indices \citep{hkk98}.  In this
paper we report on radio interferometric observations of PSRs
J0940$-$5428 and J1301$-$6305 conducted with the Australia Telescope
Compact Array (ATCA; Frater, Brooks, \& Whiteoak
1992\nocite{fbw92}). Polarization information was recorded in these
observations, and from these data we derive flux densities, spectral
indices, RMs, and polarization properties for these two pulsars and
discuss the results.

\section{Observations and Data Analysis} 

We observed PSRs J0940$-$5428 and J1301$-$6305 with the ATCA in August
1999, soon after their discovery in the Parkes Multibeam survey
\citep{mlc+01}. Each pulsar was observed in the 6D array configuration
with the 6~km antenna, which provides the highest possible spatial
resolution. The pulsars were observed simultaneously at center
frequencies of 1384 and 2496 MHz, with a bandwidth of 128 MHz at each
frequency.  Table \ref{tbl-1} presents the observing parameters and
details.  Pulsar gating was used during each observation (e.g.,
Stappers, Gaensler, \& Johnston 1999\nocite{sgj99}), which preserved
pulse phase information.  The data were reduced with the MIRIAD
software
package.\footnote{http://www.atnf.csiro.au/computing/software/miriad}
After excision of internally generated radio frequency interference
(RFI), 13 contiguous 8-MHz frequency channels remained which covered a
total bandwidth of 104 MHz at each frequency.  The data were then
further edited and flagged for RFI.  The pulse phase bins were
appropriately phase-adjusted as a function of frequency channel to
account for interstellar dispersion, after which the frequency
channels were summed.  Full Stokes parameters were recorded for each
pulse phase bin during each observation.  We measured flux densities
at both frequencies using the UVFIT and PSRPLT routines in MIRIAD.  In
both techniques, the resulting uncertainty was added in quadrature to
a 5\% contribution from the flux calibration uncertainty, taken to be
a conservative upper limit on the uncertainty for the flux calibrator,
PKS 1934$-$638.  At each frequency, a weighted mean of the two
measured flux densities was then computed (see, e.g., Crawford
(2000)\nocite{c00} for more details).

\begin{figure}
\epsscale{1.0} \centerline{\psfig{figure=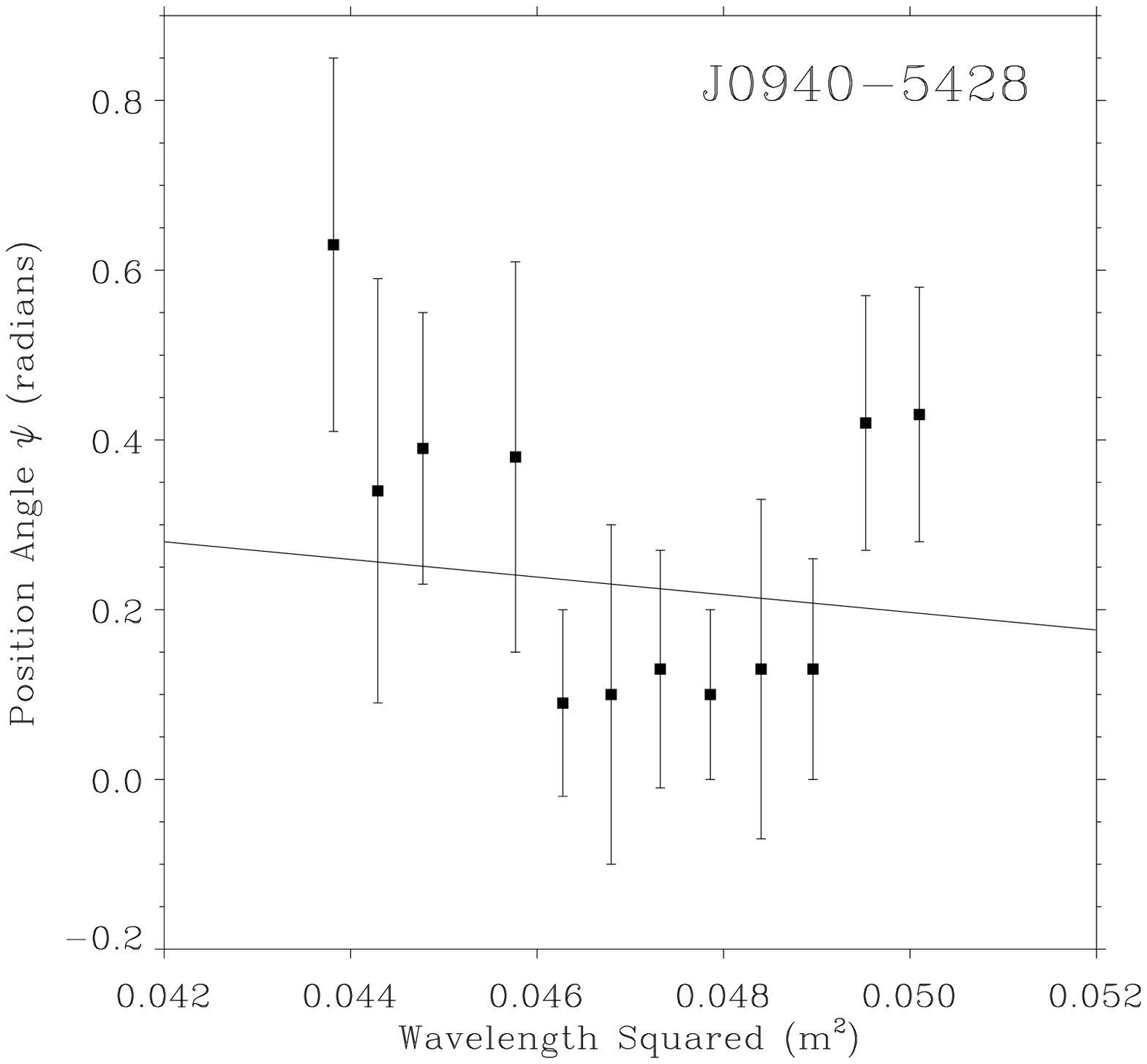,width=3in}
\psfig{figure=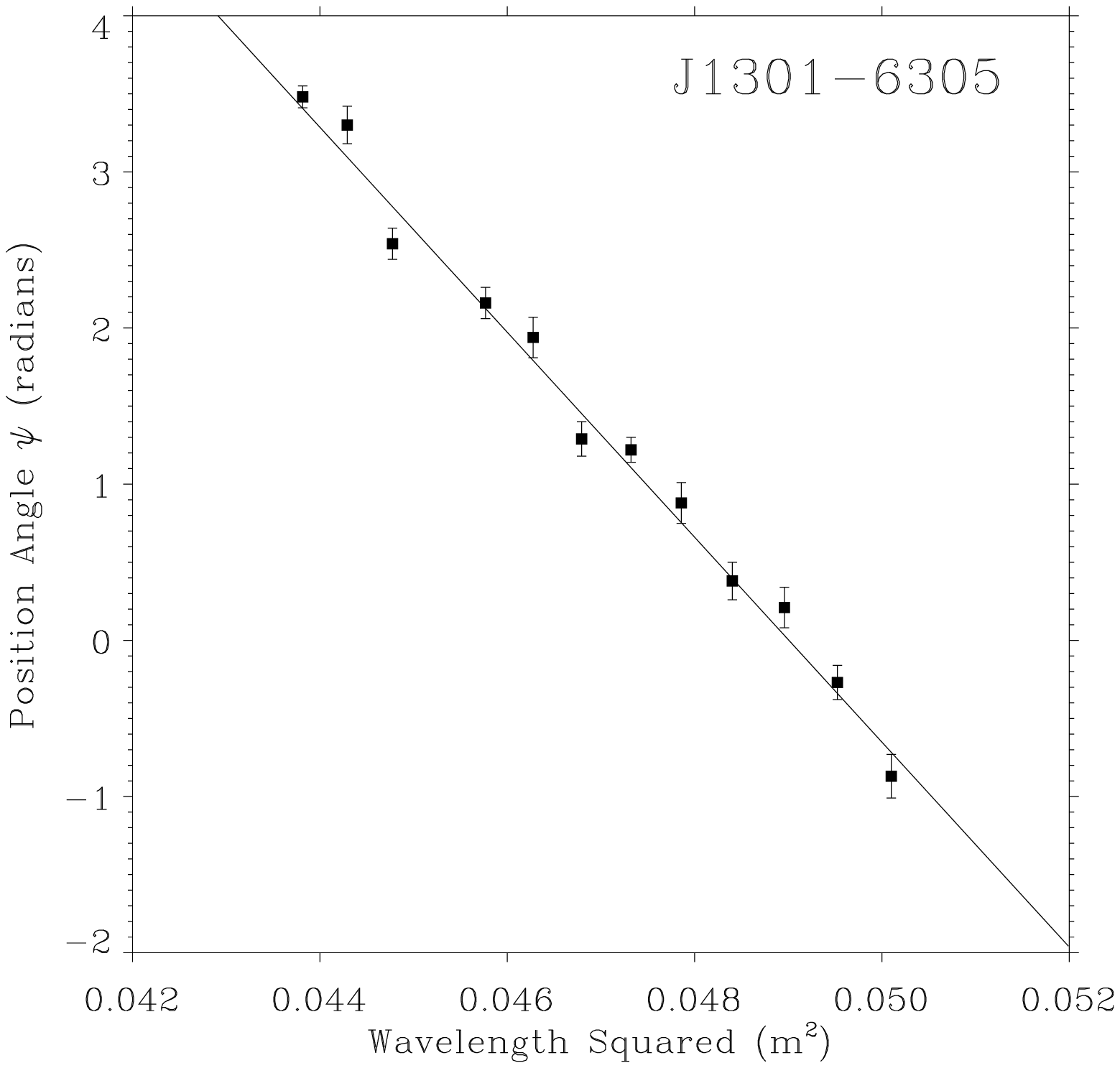,width=3in}} \figcaption{Position angle (PA)
vs. wavelength squared for PSRs J0940$-$5428 (left) and J1301$-$6305
(right).  Only the pulsed emission at 1384 MHz was used for each
plot. Several frequency channels were excised from each data set
during processing and are not used here.  The best-fit linear function
is overlaid, the slope of which is the measured RM (see Table
\ref{tbl-3}). The best-fit y-intercepts are $0.7 \pm 1.2$~rad for PSR
J0940$-$5428 and $2.8 \pm 0.7$~rad for PSR J1301$-$6305 (after an
integer number of $\pi$ phase winds are accounted for).\label{fig-1}}
\end{figure}

PSRPLT also produced Stokes parameters for each pulse phase bin. From
these, we computed the linear and circular polarization of the pulsed
emission as a fraction of the total pulsed intensity at both
frequencies. Prior to doing this, however, an RM was determined for
each pulsar using the channelized 1384-MHz data.  A position angle
(PA) $\psi$ was computed at the pulsar's location for each frequency
channel using Stokes $Q$ and $U$, according to $\psi =
1/2\arctan(U/Q)$, and a linear fit to the result was performed as a
function of wavelength, according to $\psi = \psi_{0} + {\rm
RM}\,\lambda^{2}$ (see Figure \ref{fig-1}).  A linear polarization
magnitude $L$ was computed from $L = (Q^{2} + U^{2})^{1/2}$, which was
then corrected for the positive contribution of receiver noise (see,
e.g., Manchester, Han, \& Qiao 1998\nocite{mhq98}; Crawford,
Manchester, \& Kaspi 2001\nocite{cmk01}; Crawford \& Keim
2003\nocite{ck03}). PSR J1301$-$6305 suffered from significant
bandwidth depolarization owing to its large RM, and the reported
linear polarization fraction in Table \ref{tbl-3} for this pulsar has
been corrected at both frequencies to account for this. Stokes $V$
represents circular polarization, with positive values corresponding
to left-circular polarization.

The effect of bandwidth depolarization was determined in the following
way: PAs within a bandwidth $\Delta f$ centered at a frequency $f_{0}$
are spread out in angle owing to Faraday rotation. Within the
bandwidth, the PAs span an angle $\Delta \psi_{s}$ determined by

\begin{equation}
\Delta \psi_{s} = \frac{(1.8 \times 10^{5}) \Delta f \, {\rm RM}}{f_{0}^{3}}
\end{equation}

where $\Delta \psi_{s}$ and RM are measured in rad and rad m$^{-2}$,
respectively, and the frequencies are in MHz. The sum of the PA
vectors within the bandwidth produces a net magnitude which is smaller
than the corresponding sum of aligned PAs, owing to partial
cancellation.  The measured linear polarization fraction is therefore
underestimated relative to its true value.  The ratio of these two
magnitudes, $R$, is determined by

\begin{equation} 
R = \left| \frac{\sin(\Delta \psi_{s}/2)}{(\Delta \psi_{s}/2)} \right|
\end{equation} 

where $\Delta \psi_{s}$ is again in rad (and $R \le 1)$.  The measured
linear polarization fraction can be multiplied by $1/R$ to correct for
this effect.


\section{Results and Discussion} 

\subsection{Flux Densities and Spectral Indices} 

\begin{figure}
\epsscale{1.0} \centerline{\psfig{figure=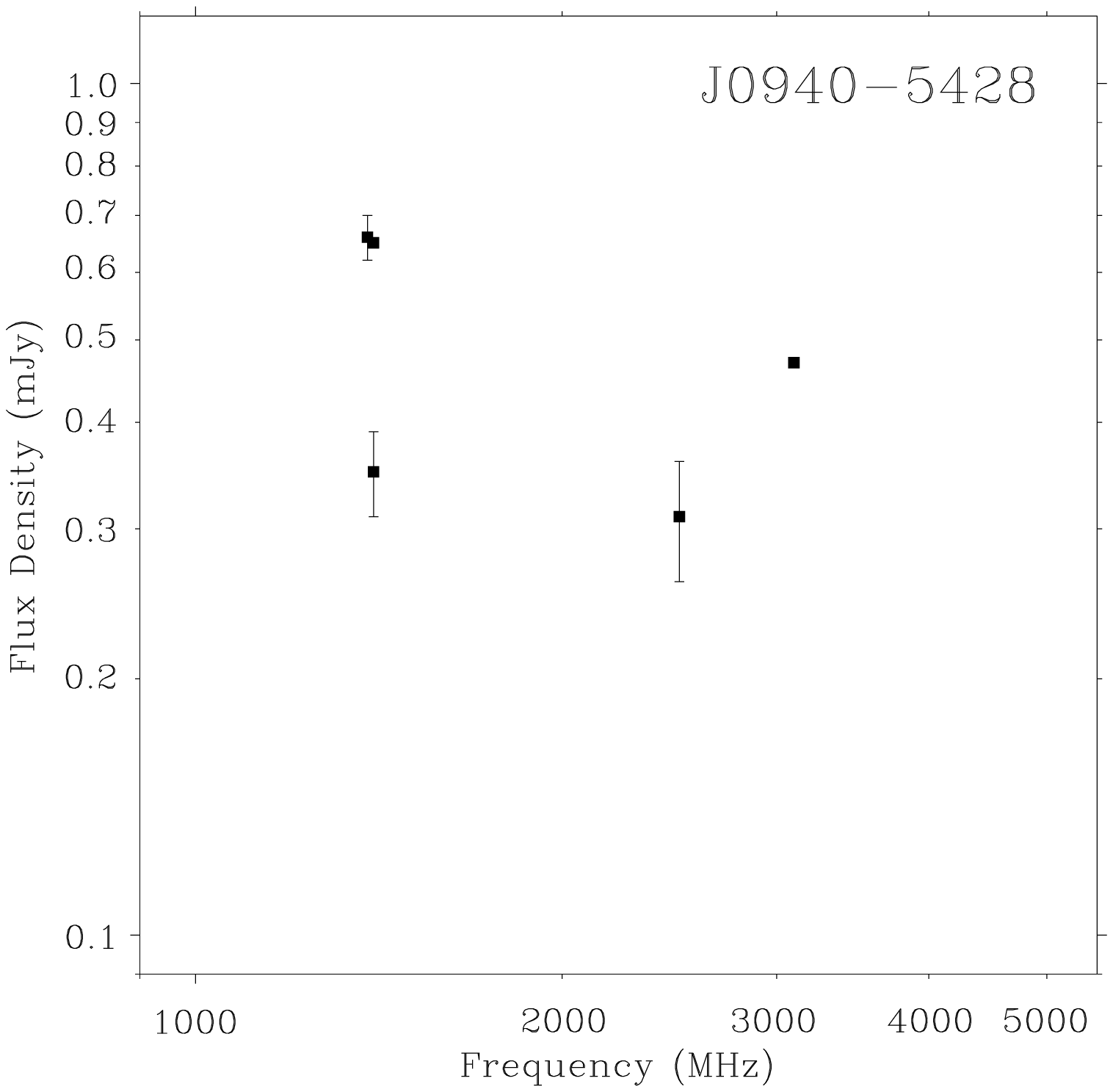,width=3in}
\psfig{figure=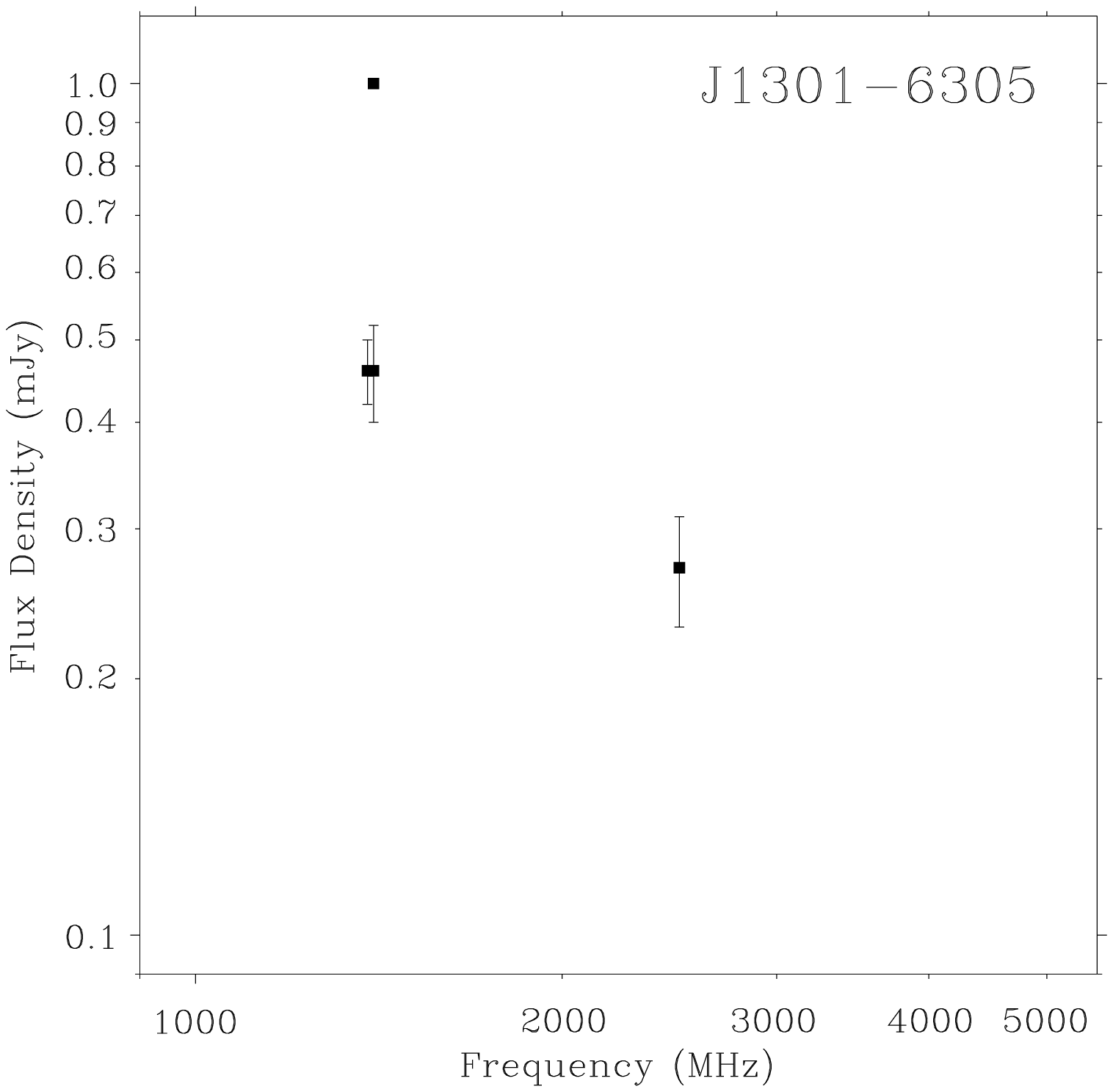,width=3in}} \figcaption{Flux density
vs. frequency for PSRs J0940$-$5428 (left) and J1301$-$6305 (right).
All measured points from Table \ref{tbl-2} are included, with
uncertainties shown where available. Both pulsars have shallow
spectral indices relative to the mean value of the known radio pulsar
population \citep{mkk+00}, but they are still within the expected
range for young, Vela-like pulsars.\label{fig-1.5}}
\end{figure}

We compared our 1384-MHz flux density measurements with those reported
for these pulsars at 1400 MHz by \citet{mlc+01} and \citet{jw06} using
single-dish observations at Parkes (see Table \ref{tbl-2}).  These
measurements, along with our measurements at 2496 MHz and a
measurement of PSR J0940$-$5428 at 3100 MHz by \citet{jw06}, are
plotted in Figure \ref{fig-1.5}.  Our 1384-MHz flux measurement for
PSR J0940$-$5428 is significantly larger than the value measured by
\citet{mlc+01}, but it is consistent with the value reported by
\citet{jw06}.  Conversely, PSR J1301$-$6305 has a measured 1384-MHz
flux density from ATCA gating which is identical to the one measured
by \citet{mlc+01}, but it is only about half of that measured by
\citet{jw06}.  The difference in these measurements may be caused by
telescope gain variations, RFI, or scintillation effects, all of which
can affect pulsar flux measurements.

Using our ATCA flux density estimates at 1384 and 2496 MHz, we
computed spectral indices for both pulsars (Table \ref{tbl-2}).  The
measured values of $\alpha = -1.3 \pm 0.3$ and $\alpha = -0.9 \pm 0.3$
(defined according to $S \sim \nu^{\alpha}$) for PSRs J0940$-$5428 and
J1301$-$6305, respectively, are both shallow relative to the mean
value of $-1.8 \pm 0.2$ for the known radio pulsar population
\citep{mkk+00}, but they are still consistent with the observed
spectral index distribution for known pulsars; both Figure 1 of
\citet{mkk+00} and the public pulsar catalog
\citep{mht+05}\footnote{http://www.atnf.csiro.au/research/pulsar/psrcat}
show that the shallow end of this distribution extends up to $\sim
0.0$.  The public pulsar catalog \citep{mht+05} lists only one pulsar
which has a measured spectral index that is positive: PSR J1740+1000
has a value of $+0.9 \pm 0.1$ measured between 0.4 and 1.4 GHz
\citep{mac+02}.  This is well above the distribution shown in Figure~1
of \citet{mkk+00} for pulsars with a single power law spectral index.
This pulsar, like the ones studied here, is fast-spinning and
energetic, with a spin period of 154 ms and a spin-down luminosity of
$2.3 \times 10^{35}$ erg s$^{-1}$. Its characteristic age of 114 kyr
places it near the age range for Vela-like pulsars.  Pulsars with
these characteristics (identified by \citet{hkk98} as the B1800$-$21
class of pulsars) can in some cases have high turnover frequencies
($\ga 1$~GHz; see, e.g., Maron et al. (2000)\nocite{mkk+00} and Kijak
et al. (2007)\nocite{kgk07}). However, \citet{mac+02} suggest that the
spectral index measurement for PSR J1740+1000 may suffer from
contamination by interstellar scintillation (refractive scintillation
in particular), which is uncorrelated between frequencies.  Thus,
although such pulsars are expected to have shallow spectral indices,
they are not expected to have positive values, and the two pulsars
studied here do indeed have shallow and negative spectral indices.

\subsection{Polarization Characteristics} 

\begin{figure}
\epsscale{1.0} \centerline{\psfig{figure=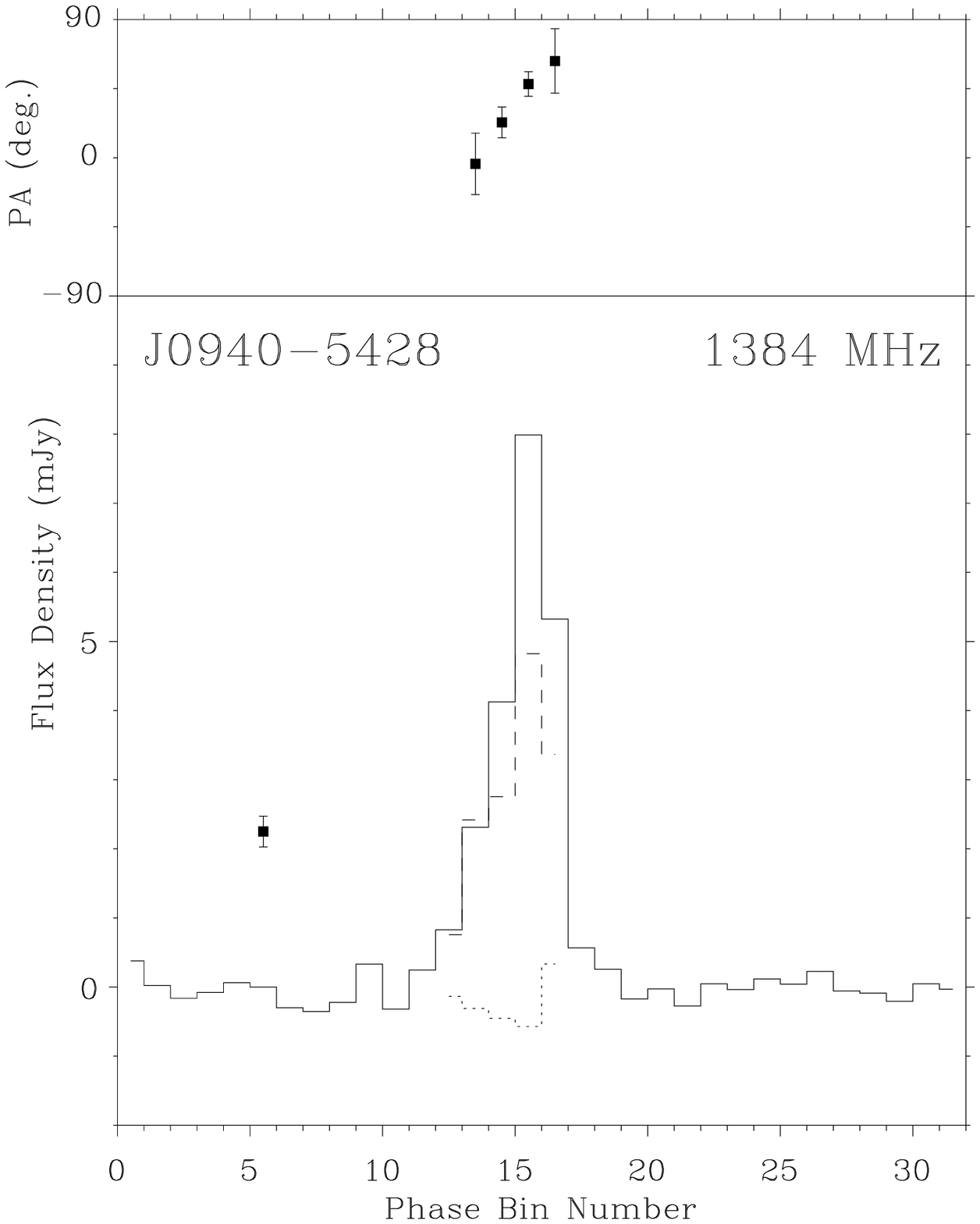,width=2.8in}
\psfig{figure=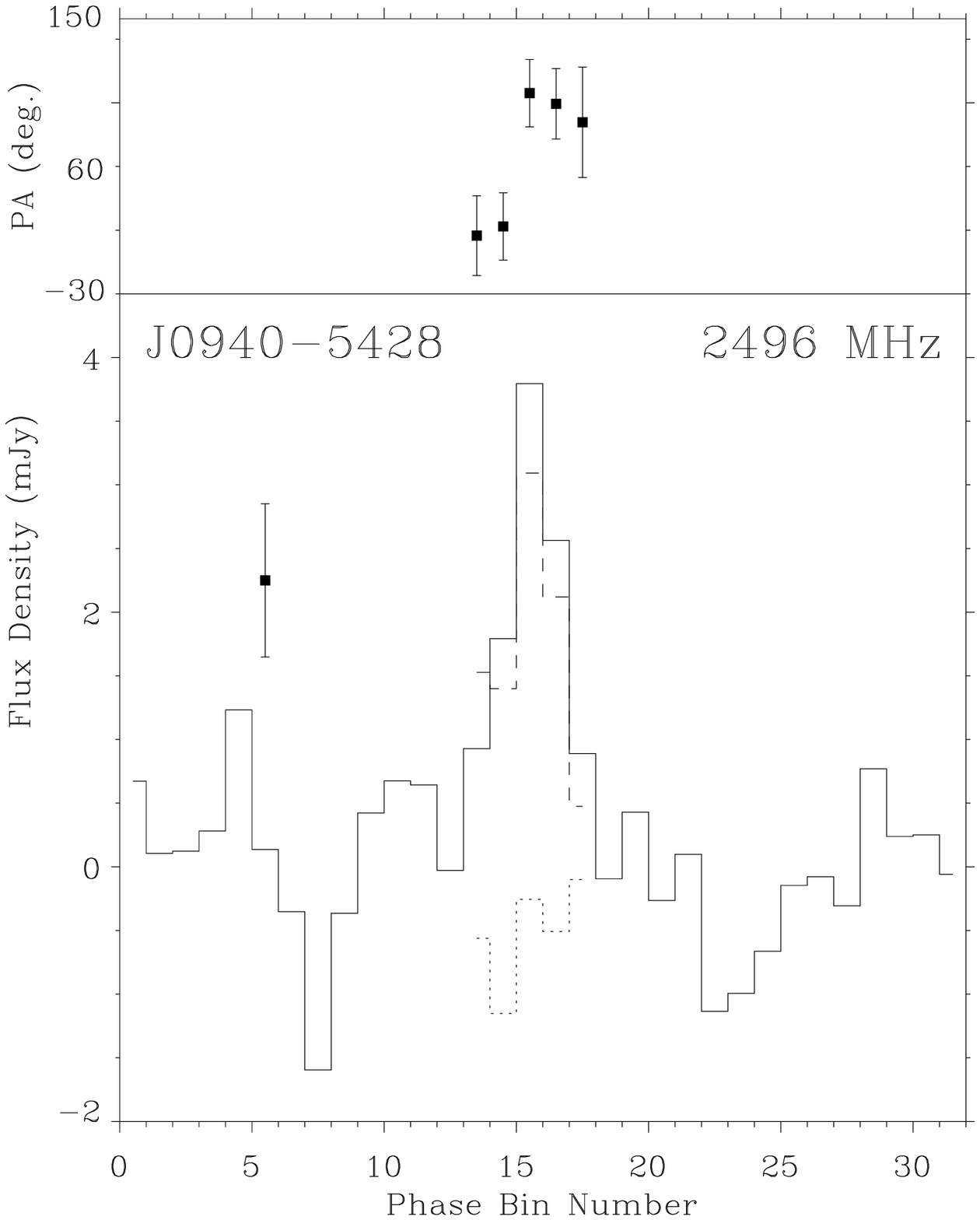,width=2.8in}}
\centerline{\psfig{figure=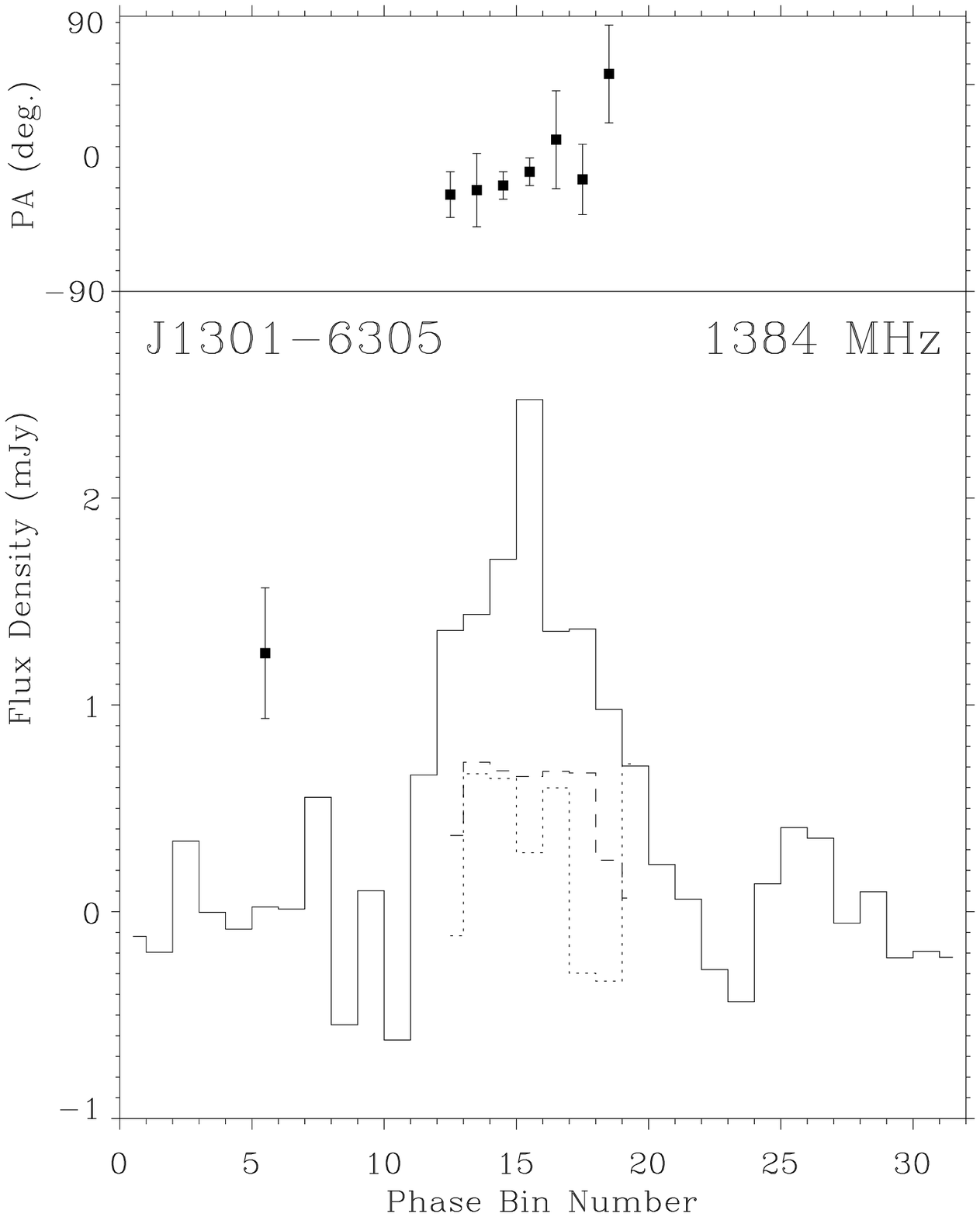,width=2.8in}
\psfig{figure=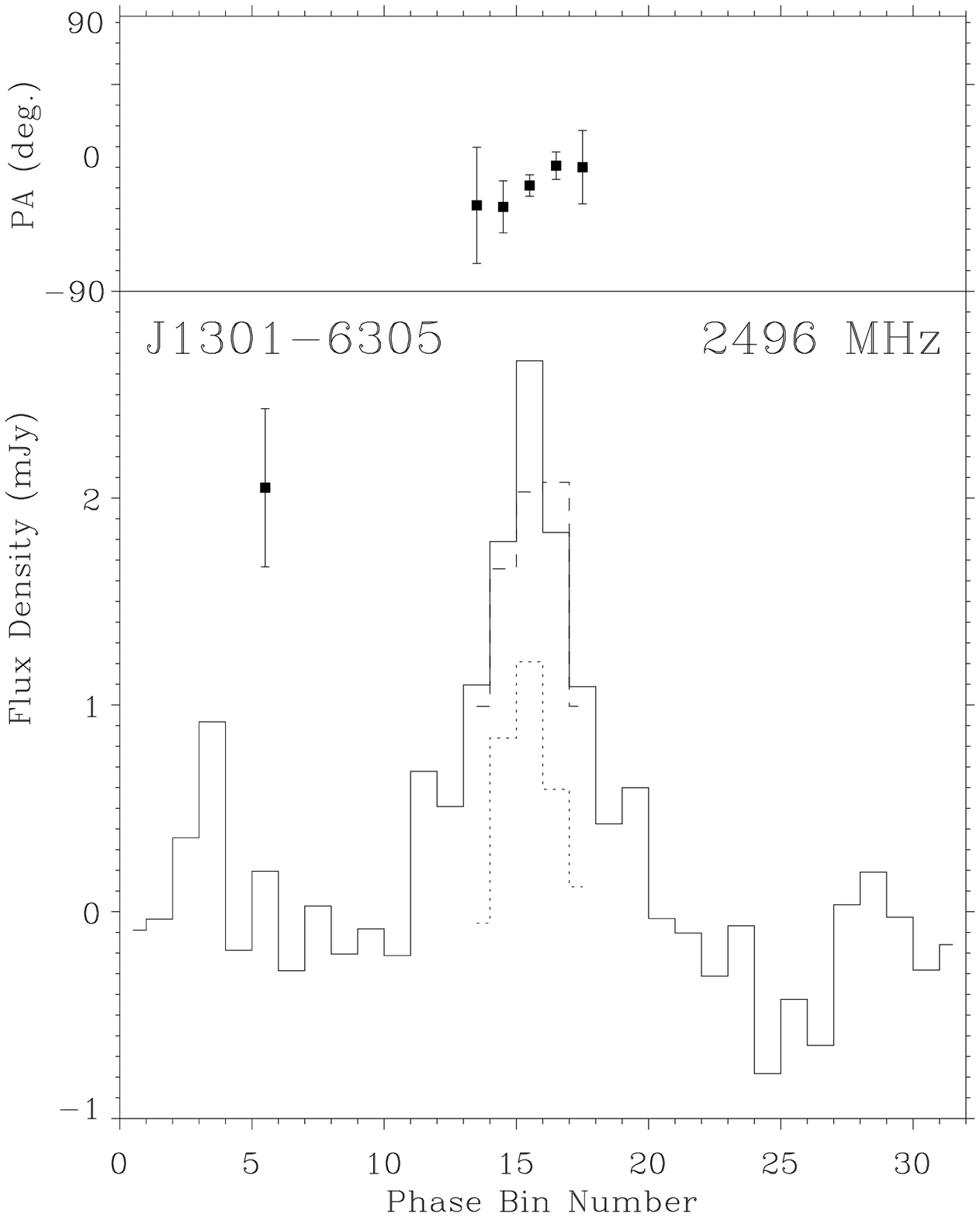,width=2.8in}} \figcaption{Polarization profiles
for PSRs J0940$-$5428 (top row) and J1301$-$6305 (bottom row) at 1384
MHz (left) and 2496 MHz (right). Each profile has 32 phase bins and
was created from data taken with the ATCA.  One full period is shown
for each profile. The off-pulse rms for each profile is indicated by
the error bar to the left of the profile, and a mean off-pulse
baseline value has been subtracted from all bins in each
profile. Solid, dashed, and dotted lines indicate total intensity,
linearly polarized intensity, and circularly polarized intensity,
respectively.  The PAs (measured from North to East) and their
uncertainties are shown above each profile bin where measurements were
possible.  Bandwidth depolarization is not accounted for in the
profiles, which, in the case of PSR J1301$-$6305 at 1384 MHz,
significantly reduces the measured linear polarization fraction
relative to its true value (see Table \ref{tbl-3}).\label{fig-2}}
\end{figure}

Both pulsars are highly polarized at 1384 and 2496 MHz.  The
phase-resolved polarization profiles and PAs constructed from the ATCA
data are shown at both frequencies in Figure \ref{fig-2}, and the
measured polarization fractions from these profiles are presented in
Table \ref{tbl-3}.

High-resolution profiles of PSR J0940$-$5428 at 1369 and 3100 MHz
presented by \citet{jw06} show that this pulsar has an asymmetric,
double-peaked profile, with the leading peak being somewhat weaker
than the trailing peak.  The separation of these peaks is $\sim
15^{\circ}$ of the pulse phase, which corresponds to less than two
bins in our ATCA profiles (Figure \ref{fig-2}). It is not surprising,
therefore, that these peaks are not resolved in our profiles. However,
a hint of the leading component may be visible at both frequencies in
Figure \ref{fig-2}, and it is highly polarized in each case. The
measured linear polarization fractions are 69\% and 86\% for the
pulsed emission at 1384 and 2496 MHz, respectively, with uncertainties
as given (see Table \ref{tbl-3}). These values are qualitatively
consistent with the polarization profiles presented by \citet{jw06},
although they do not report measured polarization fractions with which
to compare our numbers. It is clear from the profiles for PSR
J0940$-$5428 shown here and by \citet{jw06} that the pulsar remains
highly polarized across a range of frequencies.

\begin{figure}
\epsscale{1.0} \centerline{\psfig{figure=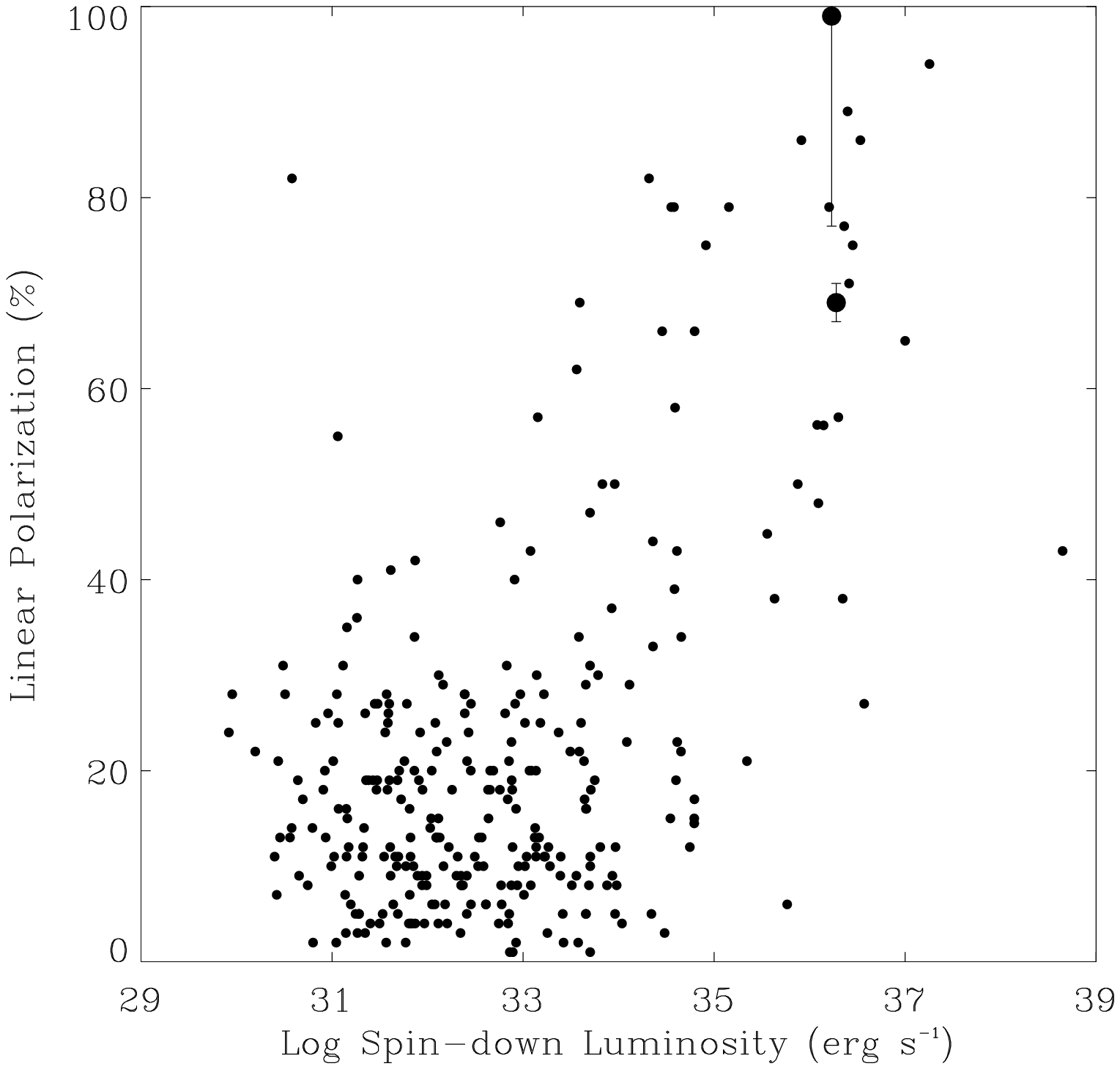,width=4in}}
\figcaption{Fractional linear polarization at 1400 MHz vs. spin-down
luminosity for pulsars with published 1400-MHz polarization
measurements.  This sample includes 278 pulsars published by
\citet{gl98} and a number of measurements made at or near 1400 MHz
published elsewhere \citep{qmlg95,mj95,cmk01,rrj01,ck03}.  This figure
is an extension of Figure~2 shown by \citet{cmk01}. The two pulsars
shown with 1$\sigma$ error bars are PSRs J0940$-$5428 and
J1301$-$6305, measured using the ATCA at 1384 MHz.  The linear
polarization fraction for PSR J1301$-$6305 has been corrected for
bandwidth depolarization and nears 100\% at this frequency.  These two
pulsars fit a previously noted correlation between spin-down
luminosity and degree of linear polarization at 1400
MHz.\label{fig-3}}
\end{figure}

PSR J1301$-$6305 has a wide profile at 1384 MHz (Figure \ref{fig-2}),
and when bandwidth depolarization is taken into consideration, the
pulsar is $\sim$~100\% polarized at both frequencies (Table
\ref{tbl-3}). A high-resolution polarization profile at 1375 MHz
presented by \citet{jw06} is consistent with the high degree of
polarization measured in our 1384-MHz data.  Our 2496 MHz data
indicates that this pulsar, like PSR J0940$-$5428, remains highly
polarized at higher frequencies. Both PSR J0940$-$5428 and PSR
J1301$-$6305 also fit a previously noted trend in which pulsars with
large spin-down luminosities ($\dot{E}$) have high linear polarization
fractions at 1400 MHz \citep[Figure \ref{fig-3}; see also][]{hkk98, cmk01}.

The phase-resolved PA data for each profile are also shown in Figure
\ref{fig-2}. The PAs are referenced with respect to celestial North,
as is the usual convention. Although variation in the PAs can be seen
as a function of pulse longitude in each case, the profile resolution
is low.  No constraints on the emission geometry are possible with
these data using the rotating-vector model of
\citet{rc69}. \citet{jw06} present an in-depth discussion about the
general properties of young pulsars in the context of polarization
measurements and identify some trends.  Apart from the correlations
previously mentioned, we also note that these two pulsars have
relatively simple pulse profile morphologies, as seen for young
pulsars more generally. The PAs also show no evidence of orthogonal
mode changes, with the possible exception of PSR J0940$-$5428 at 2496
MHz (Figure \ref{fig-2}).  There is no corresponding profile for PSR
J0940$-$5428 at this frequency in \citet{jw06}, but their 3100-MHz
profile shows no indication of such a jump.

\subsection{Rotation Measures}

An RM was measured for each pulsar using the 1384-MHz data, and these
RMs are reported in Table \ref{tbl-3}. A correction for the
ionospheric contribution to the RM was not made, but this contribution
is expected to be only a few rad m$^{-2}$, significantly smaller than
the uncertainties in the measured RMs (cf. Johnston \& Weisberg
2006\nocite{jw06} and references therein).  The measured values from
the ATCA data are consistent with the RMs reported recently by
\citet{hml+06} and \citet{jw06} using observations at Parkes.  Using
the dispersion measures (DMs) reported by \citet{mlc+01} and the
measured RMs from our ATCA observations, the mean line-of-sight
magnetic field strength was calculated for each pulsar according to
$\langle B_{\Vert} \rangle = 1.232 \, {\rm RM}/{\rm DM}$
\citep{mt77}. Here the RM and DM are in units of rad m$^{-2}$ and pc
cm$^{-3}$, respectively, and $\langle B_{\Vert} \rangle$ is in
$\mu$G. Table \ref{tbl-3} lists the calculated values of $\langle
B_{\Vert} \rangle$.

We compared published RMs of pulsars within a few degrees of PSRs
J0940$-$5428 and J1301$-$6305 with our measured RMs to see if the
values were in any way anomalous.  The 7 known pulsars that lie within
3$^{\circ}$ of the location of PSR J0940$-$5428 (at Galactic longitude
$l=277.5^{\circ}$ and latitude $b=-1.3^{\circ}$) which have measured
RMs span DM values from $\sim 100$ to $\sim 200$ pc cm$^{-3}$. The
derived values of $\langle B_{\Vert} \rangle$ for these pulsars are
scattered around zero, and $\langle B_{\Vert} \rangle$ for PSR
J0940$-$5428 (which has a DM of 136 pc cm$^{-3}$) is consistent with
this distribution.  For the 9 known pulsars lying within 3$^{\circ}$
of PSR J1301$-$6305 (located at $l=304.1^{\circ}$, $b=-0.2^{\circ}$)
that have measured RMs, there is a large range of DM values ($\sim
100$ to $\sim 700$ pc cm$^{-3}$), with the RM values trending toward
large negative values as the DM increases. This is shown in Figure~6
of \citet{hml+06}, which depicts RM as a function of both distance and
DM for pulsars in the direction of the Crux spiral arm of the Galaxy.
PSR J1301$-$6305 falls roughly in the middle of this DM range (374 pc
cm$^{-3}$) and indeed has a large negative RM.  In fact, its RM is the
largest in the negative direction of the nine pulsars, and the
inferred mean line-of-sight magnetic field strength of $\sim 2$~$\mu$G
is about twice as large as the magnitude of the next largest value in
this sample.  Still, this is not anomalously high for Galactic
values. An additional localized region of highly magnetized plasma
between us and PSR J1301$-$6305 could be enhancing the RM, but there
is no evidence for such a region in radio maps.

\section{Conclusions} 

Using pulsar-gated radio interferometric observations taken at 1384
and 2496 MHz with the ATCA, we have measured the flux densities and
polarization properties of two Vela-like pulsars, PSRs J0940$-$5428
and J1301$-$6305.  The measured spectral indices for both pulsars from
our observations are shallow but still consistent with values for the
known pulsar population \citep{mkk+00}.  The polarization properties
of the pulsed emission indicate that both pulsars are highly polarized
at both frequencies. The shallow spectral indices and the high degree
of linear polarization are both consistent with the properties of
other young, Vela-like radio pulsars, and these measurements fit a
previously established correlation between spin-down luminosity and
degree of linear polarization at 1400 MHz \citep{hkk98, cmk01}. The
RMs derived for the pulsars are consistent with measurements made at
1400 MHz using the Parkes telescope, and they yield mean line-of-sight
magnetic field strengths that are within the normal range for Galactic
values \citep{hml+06}.

\acknowledgements

We thank Elisabeth Bardenett for assistance editing and flagging the
data.  CLT was supported by a summer research grant from the Keck
Northeast Astronomy Consortium.  The ATCA is part of the Australia
Telescope, which is funded by the Commonwealth of Australia for
operation as a National Facility operated by CSIRO.

\clearpage

\clearpage

\begin{deluxetable}{cccccc}
\footnotesize
\tablecaption{ATCA Observing Parameters.\label{tbl-1}}
\tablewidth{0pt}
\tablehead{
\colhead{PSR} &
\colhead{Observing} &
\colhead{Array} &
\colhead{PKS Phase} &
\colhead{PKS Flux} &
\colhead{Integration} \\
\colhead{} & 
\colhead{Date} &
\colhead{Config.} &
\colhead{Calibrator} &
\colhead{Calibrator} &
\colhead{Time (hr)}
}
\startdata
J0940$-$5428 & 1999 Aug 18 & 6D   & 0823$-$500 & 1934$-$638 & 6.8  \\
J1301$-$6305 & 1999 Aug 22 & 6D   & 1329$-$665 & 1934$-$638 & 10.6 \\
\enddata

\tablecomments{In both observations, data were taken at center
frequencies of 1384 and 2496 MHz with 32 phase bins used in pulsar
gating. A bandwidth of 128 MHz was used for each frequency in each
observation, but this was reduced to 104 MHz after excision of some
data during processing. The reduced bandwidth consisted of 13
contiguous 8-MHz channels.}

\end{deluxetable}

\begin{deluxetable}{lcc}
\footnotesize
\tablecaption{Flux Densities and Spectral Indices.\label{tbl-2}}
\tablewidth{0pt}
\tablehead{
\colhead{PSR} &
\colhead{J0940$-$5428} &
\colhead{J1301$-$6305}
}
\startdata
1384-MHz flux density (mJy)\tablenotemark{a}      & 0.66(4)           & 0.46(4)           \\
2496-MHz flux density (mJy)\tablenotemark{a}      & 0.31(5)           & 0.27(4)           \\  
Spectral Index, $\alpha$\tablenotemark{b}         & $-$1.3(3)         & $-$0.9(3)         \\
                                                  &                   &                   \\
1400-MHz flux density (mJy)\tablenotemark{c}      & 0.35(4)           & 0.46(6)           \\
1400-MHz flux density (mJy)\tablenotemark{d}      & 0.65              & 1.00              \\
3100-MHz flux density (mJy)\tablenotemark{d}      & 0.47              & ---               \\ 
\enddata

\tablecomments{Figures in parentheses represent the $1\sigma$
uncertainty in the least significant digit quoted.}

\tablenotetext{a}{From ATCA gated data.}

\tablenotetext{b}{Spectral index $\alpha$ (defined according
to $S \sim \nu^{\alpha}$)
determined using 1384 and 2496 MHz ATCA flux densities.}

\tablenotetext{c}{From Parkes timing observations \citep{mlc+01}.}

\tablenotetext{d}{From Parkes polarization observations \citep{jw06}.}

\end{deluxetable}

\begin{deluxetable}{lcccc}
\footnotesize
\tablecaption{Polarization Characteristics.\label{tbl-3}}
\tablewidth{0pt}
\tablehead{
\colhead{PSR} &
\multicolumn{2}{c}{J0940$-$5428} &
\multicolumn{2}{c}{J1301$-$6305} 
}
\startdata
Frequency (MHz)                                            & 1384             & 2496         & 1384                          & 2496          \\ \\
$\langle L \rangle / S$ (\%)\tablenotemark{a}              & $69 \pm 2$       & $86 \pm 14$  & $99  \pm 22$\tablenotemark{f} & $94 \pm 10$\tablenotemark{f}  \\
$\langle V \rangle / S$ (\%)\tablenotemark{b}              & $-6 \pm 2$       & $-26 \pm 14$ & $+19 \pm 8$                   & $+32 \pm 10$  \\
$\langle \vert V \vert \rangle / S$ (\%)\tablenotemark{c}  & $ 9 \pm 2$       & $26 \pm 14$  & $32  \pm 8$                   & $33  \pm 10$  \\ \\
RM (rad m$^{-2}$)\tablenotemark{d}                         & $-10 \pm 24$     & ---          & $-631 \pm 15$                 & ---         \\
$\langle B_\Vert \rangle$ ($\mu$G)\tablenotemark{e}        & $-0.09 \pm 0.22$ & ---          & $-2.08 \pm 0.05$              & ---         \\
\enddata

\tablecomments{~Listed uncertainties are at the 1$\sigma$ level,
and all percentages have been rounded to the nearest whole number.}

\tablenotetext{a}{Fractional on-pulse linear polarization.}

\tablenotetext{b}{Fractional on-pulse circular polarization. Positive
values correspond to left circular polarization.}

\tablenotetext{c}{Fractional on-pulse absolute circular polarization.}

\tablenotetext{d}{Rotation measure derived from the 1384-MHz ATCA data.}

\tablenotetext{e}{Mean line-of-sight magnetic field strength. Negative
values correspond to field lines pointing away from the observer.}

\tablenotetext{f}{Values have been corrected for bandwidth
depolarization using the measured RM and a bandwidth of 104 MHz. 
Values measured for PSR
J1301$-$6305 prior to correction 
were $35 \pm 8$\% at 1384 MHz and $91 \pm 10$\% at 2496
MHz.}

\end{deluxetable}

\end{document}